\documentclass[reprint, twocolumn,pra,amsmath,amssymb,longbibliography, 10pt, aps, superscriptaddress]{revtex4-1}
\usepackage{graphicx}
\usepackage{amsmath,latexsym,tabularx}
\usepackage{xcolor}
\usepackage[%
  colorlinks=true,
  urlcolor=blue,
  linkcolor=blue,
  citecolor=blue
]{hyperref}
\usepackage{booktabs}

\newcommand{\eqnRef}[1]{\mbox{Eq.\,(\ref{#1})}}

\newcommand{\figRef}[1]{\mbox{Fig.\,\ref{#1}}}

\newcommand{\affilLL}[0]{Lincoln Laboratory, Massachusetts Institute of Technology, Lexington, Massachusetts 02421, USA}
\newcommand{\affilMITPhys}{Department of Physics, Massachusetts Institute of Technology, Cambridge, Massachusetts 02139, USA}
\newcommand{\affilMITEECS}{Department of Electrical Engineering and Computer Science, Massachusetts Institute of Technology, Cambridge, Massachusetts 02139, USA}
\newcommand{\affilMITRLE}{Research Laboratory of Electronics, Massachusetts Institute of Technology, Cambridge, Massachusetts 02139, USA}
\newcommand{\unit}[1]{\, \mathrm{#1}}

\begin{document}

\title{Chip-integrated voltage sources for control of trapped ions}

\author{J. Stuart}
\email[]{jstuart@mit.edu}
\affiliation{\affilMITRLE}
\affiliation{\affilMITPhys}
\affiliation{\affilLL}

\author{R. Panock}
\affiliation{\affilLL}

\author{C. D. Bruzewicz}
\affiliation{\affilLL}

\author{J. A. Sedlacek}
\thanks{Present address: Honeywell, Golden Valley, MN}
\affiliation{\affilLL}

\author{R. McConnell}
\affiliation{\affilLL}

\author{I. L. Chuang}
\affiliation{\affilMITRLE}
\affiliation{\affilMITPhys}
\affiliation{\affilMITEECS}

\author{J. M. Sage}
\affiliation{\affilLL}
\affiliation{\affilMITRLE}

\author{J. Chiaverini}
\affiliation{\affilLL}

\date{\today}


\begin{abstract}
Trapped-ion quantum information processors offer many advantages for achieving high-fidelity operations on a large number of qubits, but current experiments require bulky external equipment for classical and quantum control of many ions. We demonstrate the cryogenic operation of an ion-trap that incorporates monolithically-integrated high-voltage CMOS electronics ($\pm 8\unit{V}$ full swing) to generate surface-electrode control potentials without the need for external, analog voltage sources. A serial bus programs an array of 16 digital-to-analog converters (DACs) within a single chip that apply voltages to segmented electrodes on the chip to control ion motion. Additionally, we present the incorporation of an integrated circuit that uses an analog switch to reduce voltage noise on trap electrodes due to the integrated amplifiers by over $50\unit{dB}$. We verify the function of our integrated electronics by performing diagnostics with trapped ions and find noise and speed performance similar to those we observe using external control elements.
     
\end{abstract}

\maketitle



\section{Introduction}

Quantum computers and simulators based on trapped atomic ions have great potential to allow for the execution of complex algorithms, but to date, experiments have been limited to tens of ion qubits.  Increasing the size of linear chains of ions presents experimental challenges including optical addressing and control of individual ions, arbitrary re-ordering of ion positions during complex protocols, and limited lifetimes of long chains \cite{Kaufmann2017, Pagano2018}.  Alternatively, array-based designs which naturally solve these issues have been discussed recently \cite{Kielpinski2002, Lekitsch2017}.  The array architecture has a clearer path toward increasingly-complex designs and might also benefit from modern semiconductor fabrication techniques.  These architectures will require integration of control elements into the vacuum chamber to reduce the number of required interconnects.  Off-the-shelf electronics have been modified and attached to trap hardware in-vacuum \cite{Eltony2013, Guise2014}, but as array architectures expand beyond tens of control zones, the required circuitry for electrode control within a zone will occupy significantly more area than the footprint of an array site.  Integrating devices into the trap array itself could potentially solve this scaling problem, but this design will require a tradeoff among device area, power, speed, and noise.  Beyond quantum information processing, other chip-based technologies, like miniature atomic clocks \cite{Delehaye2018} or sensors \cite{Salim2011} could also benefit from these highly-integrated, low-power control systems.

In this work, we advance the integration of control voltages to the microscopic level by presenting a design for a voltage source, with performance similar to modern external sources, that is incorporated into the substrate beneath the electrodes of an ion trap.  Surface electrode ion traps are typically fabricated in custom lithographic processes that involve deposition and patterning of a few metal layers \cite{Seidelin2006, Stick2006, Amini2010, Hughes2011, Mielenz2016, Bruzewicz2016}. Recently, it has been shown that planar ion traps can also be manufactured in commercial CMOS processes, which offer improved design reproducibility and can reduce fabrication costs of more complex devices \cite{Mehta2014}. Here we extend the foundry-produced ion trap design to include voltage sources integrated into the semiconductor substrate.  Adding active electronics in such close proximity to the trap presents a challenge since ions are sensitive to both the static and fluctuating electric fields generated within the integrated circuit.  We carefully characterize the noise and speed of these chip-integrated devices for quantum information applications; the architecture may also be of general interest in deployable systems, due to the high repeatability and functionality afforded by CMOS technology. Explorations such as this, which evaluate the performance of quantum systems with advanced classical control, are key to assessing the utility of modern control technology for future quantum sensors, simulators, and computers.


\section{High-Voltage Digital-to-Analog Converter Design}

\begin{figure*}[t b !]
\includegraphics[width = 0.95 \textwidth]{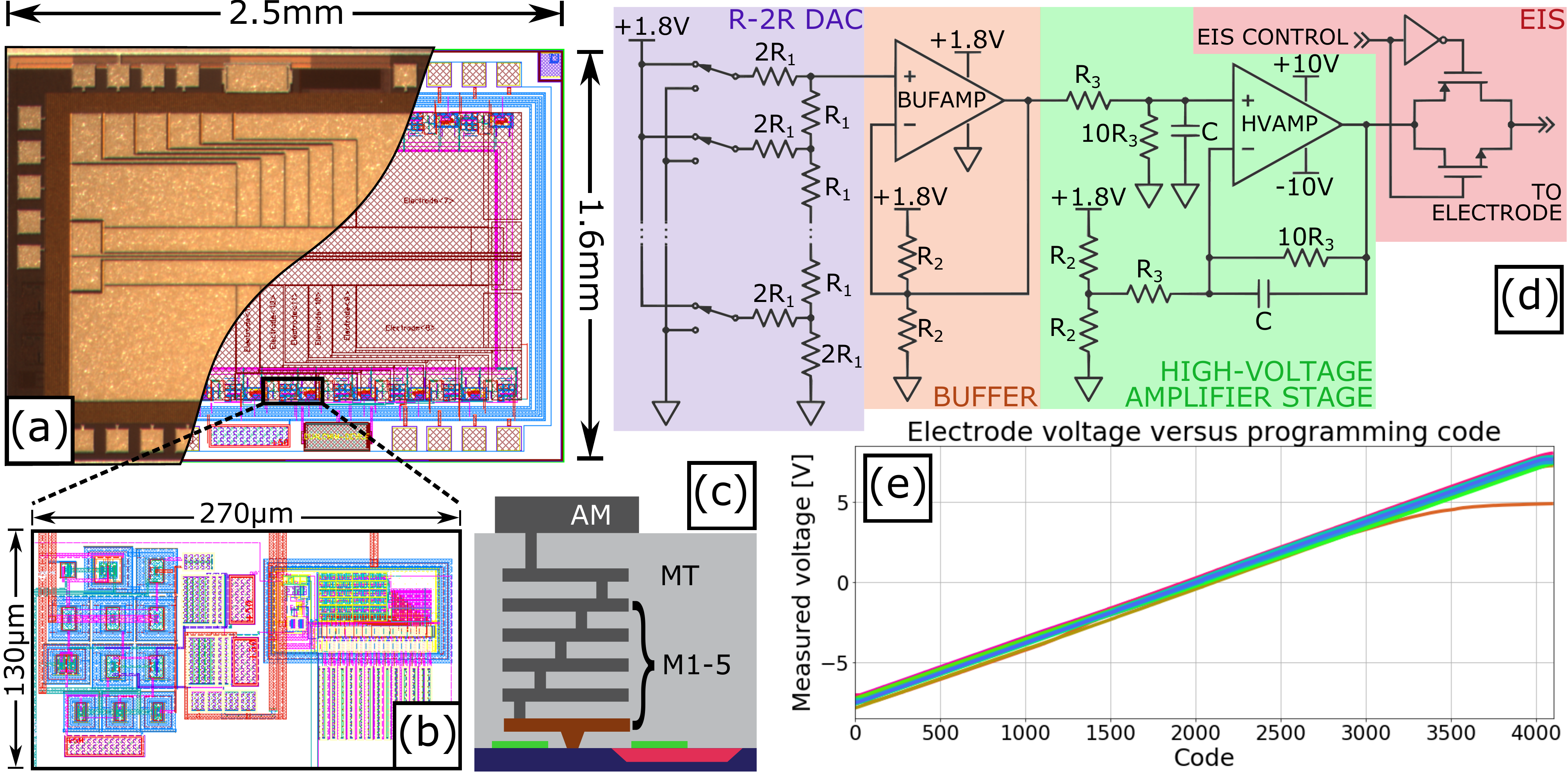}
\caption{[Color online] (a) Micrograph of the top metal layer of the chip.  The small squares around the edge of the chip are wire bond pads for connecting power (four redundant groups of 4) and digital control (one group of 5). The two largest parallel electrodes in the center of the trap are the RF electrodes, which provide ponderomotive radial confinement and are controlled by an external source.  The other 14 electrodes at the top and bottom of the chip plus 2 additional electrodes in the middle of the trap are the DC electrodes, which provide axial confinement and are controlled by the integrated voltage sources.  The cutaway reveals the layout of the circuit in the internal layers.  The different colors in the drawing represent different signal planes.  (b) Layout of one complete DAC block shown in the schematic in (d).  There are 16 such blocks on the chip.  The integrated devices fit beneath the trap electrodes and take up $15\%$ of the total device area. The blue squares on the left side of the diagram are the high-voltage FETs; each is approximately $27\unit{\mu m}\times29\unit{\mu m}$. (c) Layer stack-up for the CMHV7SF $180\unit{nm}$ node (not to scale). The total thickness of the lithographic layers is about $15\unit{\mu m}$. Trap electrodes are defined in AM. The lowest metal layer (M1) is made of copper, unlike the other (aluminum) metal layers, and has a higher current capacity.  M1 is used for routing power around the edge of the chip.  Active devices (e.g. DACs and amplifiers) are patterned in the silicon layers at the bottom of the diagram. (d) Schematic illustrating the four stages of each integrated circuit block.  The R-2R DAC converts a 12-bit digital code into an analog voltage from $0\unit{V}$ to $1.8\unit{V}$.  A SPI bus (not shown) programs the state of the switches in the R-2R. A unity-gain buffer (BUFAMP) follows the DAC and isolates the resistor ladder from the rest of the circuit.  The high-voltage amplifier (HVAMP) circuit has a gain of approximately $9\times$ and also centers the output range at $0\unit{V}$. The output electrode isolation switch (EIS) isolates the trap electrode from noise in the amplifier. The values of the components in the circuit are $\mathrm{R}_1=70\unit{k\Omega}$, $\mathrm{R}_2=2\unit{k\Omega}$, $\mathrm{R}_3=8\unit{k\Omega}$ and $\mathrm{C}=200\unit{fF}$. (e) Output voltage as a function of the 12-bit digital code sent to the DAC, for each of the 16 electrodes, measured while the trap is installed in the cryostat.  There is some variability in the voltage output between electrodes on the order of $1\unit{V}$, and one of the electrodes has significant nonlinearity at $3\unit{V}$ and above.  Both of these effects are calibrated \textit{in situ} by using wire bonds to monitor the electrode voltages.}
\label{fig:trapPic}
\end{figure*}

Voltages are derived from an R-2R resistor ladder digital-to-analog converter (DAC) \cite{Horowitz2015} with 12-bits of resolution and an output range of approximately $\pm8\unit{V}$ (see \figRef{fig:trapPic}).  The DAC accepts a 12-bit code word that it translates to an analog voltage on its output (see example data in \figRef{fig:trapPic}e).  We program the DACs using an integrated serial peripheral interface (SPI) bus, controlled by an external field-programmable gate array (FPGA) running custom firmware.  The ion trap has 16 control electrodes, with one DAC to independently bias each one.  Data is serially daisy chained through the SPI bus, for a total of $16\times12=192$ bits of voltage data.  In order to update the voltage on any or all electrodes, the full 192-bit string is sent to the chip.

The devices were fabricated as part of a multi-project wafer through MOSIS in the Global Foundries CMHV7SF $180\unit{nm}$ node.  We chose this process for its compatibility with high voltages (up to $\pm10\unit{V}$) which are necessary to achieve ion trapping frequencies around $1\unit{MHz}$ for $\mathrm{Ca}^+$ in typical ion traps.  The process includes 7 metal layers (\figRef{fig:trapPic}c); the topmost metal layer (AM), consisting of $4\unit{\mu m}$-thick aluminum, is used to define the trap electrode geometry, which is designed to trap ions at a height of $50\unit{\mu m}$.  The next lower layer (MT) forms a ground plane to isolate the ion from digital and power supply noise in the internal layers of the DAC, to isolate the DAC from pickup due to the large RF voltage on the top of the chip, and to shield the ion from laser-induced effects of carrier generation in the silicon substrate \cite{Mehta2014}.  The remaining metal layers route digital and analog signals between the CMOS transistors defined in the silicon substrate and the serial inputs and control electrodes on the top metal.  Since each high-voltage transistor is comparatively large (${\sim}800\unit{\mu m}^2$, about $2\%$ of the total DAC block area in \figRef{fig:trapPic}b), we try to use as few as possible. Hence, the first two stages of the circuit, depicted in \figRef{fig:trapPic}d, are formed using only smaller, low-voltage CMOS devices.  The number of transistors in the high-voltage amplifier (HVAMP) is also minimized, which reduces input impedance and limits the output range to $\pm8\unit{V}$ instead of the full-scale $\pm10\unit{V}$ set by the power supplies.  Due to the low input impedance, a low-voltage buffer amplifier (BUFAMP) was added between the R-2R DAC and the high-voltage amplifier stage so that current in the DAC does not leak into the HVAMP.


\section{DAC Characterization}

In our apparatus, we cool the ion trap to cryogenic temperatures (${\sim}4\unit{K}$), since this has been shown to reduce electric field noise from the trap surface and also removes the need to bake the vacuum chamber for several days to reach ultra-high vacuum \cite{Labaziewicz2008, Bruzewicz2015}.  Cryogenic operation comes with challenges in the integrated circuit performance, since the specifications provided by the CMOS foundry do not characterize behavior at temperatures below $220\unit{K}$. Thus we sacrifice some of the reproducibility usually afforded by CMOS and must ensure basic operation at our chosen temperature.  We first test the DACs on the bench by connecting the trap chip to an interposing circuit board.  The test assembly is immersed in liquid nitrogen or helium while we perform a sweep of programming codes and measure the resultant voltage on the output.  We find that some DAC channels lose the requisite linearity between programming code and output voltage as they cool to cryogenic temperature, as in \figRef{fig:trapPic}e.  For this reason, we add wire bonds to the edge of each electrode so that DAC voltages may be monitored and calibrated while conducting experiments with ions.  In future designs, we plan to add an internal multiplexer for routing the voltages to a single diagnostic pad, obviating the need for the extra bonds.

\begin{figure}[t b !]
\includegraphics[width = 0.95 \columnwidth]{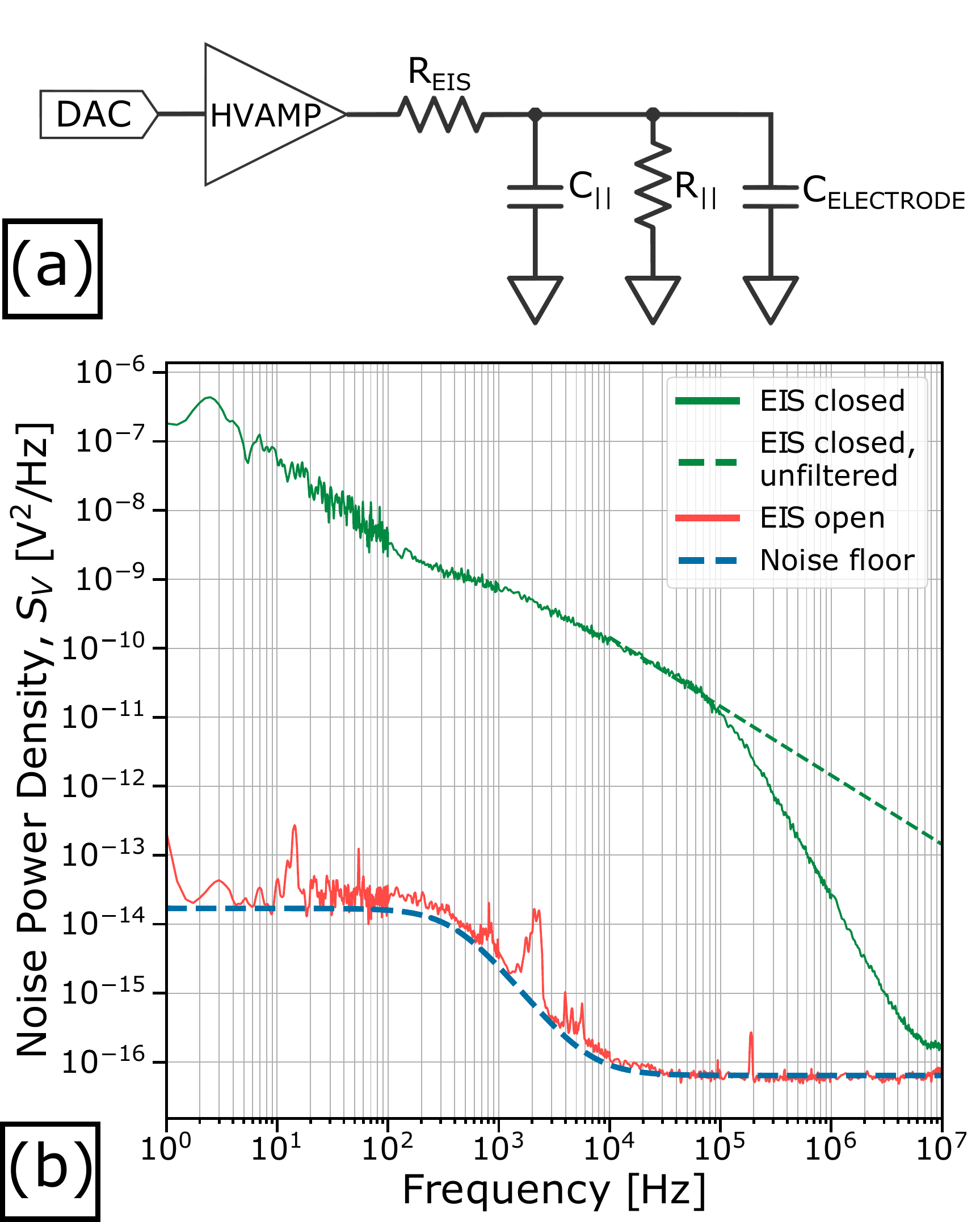}
\caption{[Color online] (a) Simplified schematic of the measurement configuration.  The capacitance of the electrode, $\mathrm{C}_\mathrm{ELECTRODE}$, set by the trap geometry and the thickness of the internal layers, is ${\sim}1\unit{pF}$.  The parallel capacitance $\mathrm{C}_\mathrm{||}$ is the sum of any external capacitance added for filtering and any capacitance from signal wires or instruments.  The parallel resistance $\mathrm{R}_\mathrm{||}$ is only present when the electrode noise is being measured by the spectrum analyzer.  (b) Measured noise power density (square of voltage noise spectral density) on one of the trap electrodes.  The voltage noise spectrum is measured using an HP89410A FFT analyzer while the trap is immersed in liquid nitrogen ($T=77\unit{K}$). As discussed in the text (see Section IV), the temperature of the trap reaches approximately $50\unit{K}$ on the cryostat, so the measured noise at $77\unit{K}$ should be representative of the noise \textit{in situ}. The noise measurement with the EIS open is limited by the noise floor of the spectrum analyzer (dashed blue line).  The shape of the EIS-open curve is due to the Johnson noise of the $1\unit{M\Omega}$ input resistance of the analyzer ($130\unit{nV}/\sqrt{\unit{Hz}}$) rolling off to the noise floor ($8\unit{nV}/\sqrt{\unit{Hz}}$) due to a pole created by the analyzer's input resistance ($\mathrm{R}_\mathrm{||}=1\unit{M\Omega}$) and parallel input capacitance ($\mathrm{C}_\mathrm{||}=400\unit{pF}$).  This parallel capacitance also causes the measured noise to roll off when the EIS is closed, due to the pole formed at $120\unit{kHz}$ with the finite ``on'' resistance of the FETs in the EIS ($\mathrm{R}_\mathrm{EIS}=3.3\unit{k\Omega}$).  The unfiltered noise is extrapolated from data taken at frequencies below this pole, assuming a $1/f$ frequency scaling (dashed green line).  At the ion's axial frequency of $1.5\unit{MHz}$, the unfiltered noise from the DACs would be $0.98\unit{\mu V}/\sqrt{\unit{Hz}}$.}
\label{fig:benchNoise}
\end{figure}

We can correct for DC voltage error in the DAC output by simply adding an offset to the programmed value. Noise at the output of the high-voltage amplifier, which cannot be calibrated away, presents a more fundamental concern for quantum logic gates based on shared motional states of chains of ions \cite{Haffner2008}.  We characterize the noise of the integrated voltage source in an external test fixture (\figRef{fig:benchNoise}b, green trace).  The extrapolated, unfiltered noise at $1\unit{MHz}$ (dashed green trace), ${\sim}1\unit{\mu V}/\sqrt{\unit{Hz}}$, is a few orders of magnitude above state-of-the-art integrated circuits with similar update speed, which have voltage noise as low as $8$--$30\unit{nV}/\sqrt{\unit{Hz}}$, though in a larger footprint \footnote{See for instance the Analog Devices AD5790 or the Texas Instruments TI-DAC8881.}.  Noise on external voltage sources can be mitigated using commercial, rack-mountable low-pass filter arrays with several orders of magnitude of suppression at relevant ion frequencies.  On the integrated device, filtering options are limited since we cannot define arbitrary inductors and capacitors on-chip due to the small size of the circuit.  Standard passive filtering also permanently reduces the overall bandwidth of the system and places a limit on the speed of ion transport operations.  To solve these problems, we have designed a switchable filter by placing a complementary pair of high-voltage field-effect transistors (FETs) between the output of the amplifier and the trap electrode, shown in the schematic of \figRef{fig:trapPic}d.  For clarity, we will refer to this device as the electrode isolation switch(es) or EIS.

The EIS approximates the function of a mechanical relay, in which no current flows and the noise voltage is perfectly attenuated when the EIS is open.  More precisely, the EIS is a voltage-variable resistance which is approximately $3.3\unit{k\Omega}$ when closed and ${\sim}\unit{T\Omega}$ when open.  On the bench, we measure a noise power attenuation of up to six orders of magnitude after opening the EIS.  This measurement is limited by the noise floor of our instruments (see \figRef{fig:benchNoise}b), so the actual isolation may be much higher.  Since trap electrodes behave as capacitors with very low leakage, we can quickly manipulate trap voltages with the EIS closed and then effectively disconnect the electrodes from the amplifier noise by opening the EIS, while the trapping voltages are maintained.  In more complex designs, the EIS could also be used for switching between multiple integrated voltage sources, which has been shown to be useful in experiments with fast ion transport \cite{Alonso2013, Alonso2016}.


\section{Operation with Trapped Ions}

During experiments with ions, we operate the trap in a cryogenic vacuum apparatus (described in \cite{Bruzewicz2016}).  The ion trap is thermally anchored to the cold head of a cryostat via a ceramic pin grid array (CPGA) and a custom printed circuit board.  An RF signal with an amplitude of $45\unit{V}$ at $45\unit{MHz}$ provides ponderomotive confinement of ions in two (radial) dimensions.  The DC voltages programmed by the DACs define the trapping potential along the third (axial) direction.  Neutral $^{40}\mathrm{Ca}$ atoms are introduced into the region above the trap by accelerating a pre-cooled cloud of atoms in a 2D MOT with a resonant ``push'' beam \cite{Sage2012}. We trap $^{40}\mathrm{Ca}^+$ after two-photon photoionization via excitation at $423\unit{nm}$ and $375\unit{nm}$.  Ions are Doppler cooled with light at $397\unit{nm}$ to about $1\unit{mK}$.  A repumping laser at $866\unit{nm}$ prevents shelving in the long-lived $\mathrm{D}_{3/2}$ state (see e.g. \cite{Bruzewicz2017} for an illustration of the electronic energy levels of $^{40}\mathrm{Ca}^+$).  A stable and narrow laser at $729\unit{nm}$ addresses the quadrupole $\mathrm{S}_{1/2}\leftrightarrow\mathrm{D}_{5/2}$ transition, which we use to read out the ion's motional state and perform resolved motional-sideband cooling close to the ground state of axial motion (${\sim}20\unit{\mu K}$) \cite{Leibfried2003}.

After applying the relatively high-voltage RF potential to the chip, we confirm that the DACs are still operational by sweeping through all possible programming codes while monitoring the voltage on each electrode via the monitor wire bonds.  We find no difference in  performance, even when applying RF voltages of up to $80\unit{V}$.  We have programmed the DAC voltages to demonstrate trapping of calcium ions at axial frequencies from $800\unit{kHz}$ up to $1.6\unit{MHz}$.  We have also performed rudimentary ion transport over a distance of $80\unit{\mu m}$ at $2\unit{mm}/\unit{s}$ by writing a continuous stream of voltages to trap electrodes at a SPI clock speed of $200\unit{kHz}$. Ion motion is verified by watching ion fluorescence on an electron-multiplying CCD camera while the ion is repeatedly shuttled back and forth by varying the DAC voltages with the EIS closed.  Ions were shuttled over hundreds of cycles without loss.  When DAC voltages are fixed, trap lifetimes are similar to those we observe in standard planar traps controlled by external voltages sources (${>}30$ seconds without cooling lasers, and several hours with Doppler cooling).  We finely adjust the internal voltages to trim out stray electric fields, due primarily to charging from photoionization lasers \cite{Harlander2010, Shannon2011}, in order to eliminate excess micromotion induced by the RF potential \cite{Sedlacek2018}.  We verify that trap voltages do not drift when the EIS is open by monitoring the trapped ion's position and trapping frequency and find that we can leave the EIS open for several minutes with no noticeable effect to the potential at the ion's location.

\begin{figure}[t b !]
\includegraphics[width = 0.95 \columnwidth]{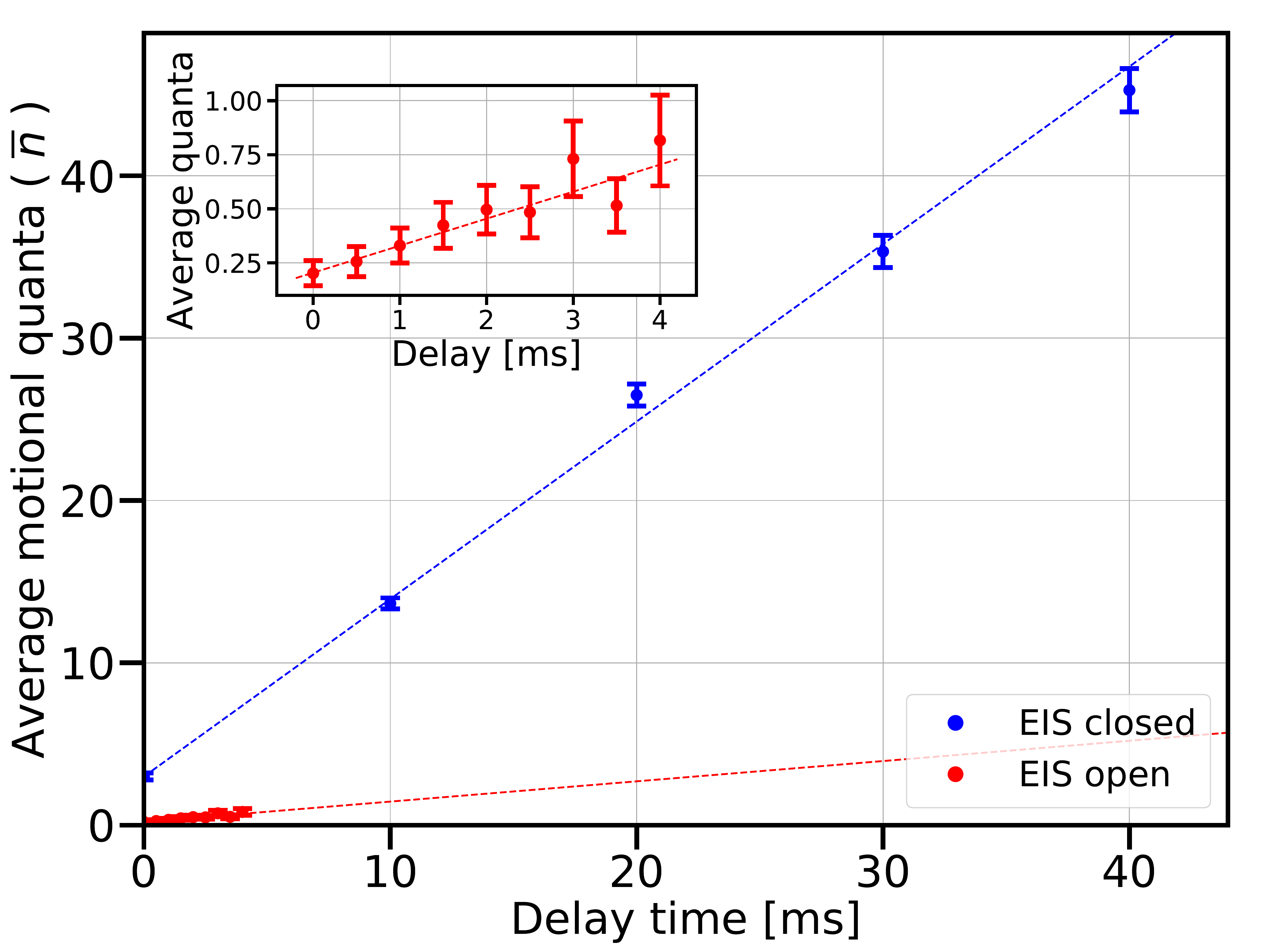}
\caption{[Color online] Axial heating rates of a single $^{40}\mathrm{Ca}^+$ ion with a trapping frequency of $1.5\unit{MHz}$.  The heating rate measured with the EIS open [closed] is $120\pm30\unit{quanta}/\unit{s}$ [$1090\pm20\unit{quanta}/\unit{s}$], which corresponds to a voltage noise of $6.0\pm0.8\unit{nV}/\sqrt{\unit{Hz}}$ [$18.2\pm0.2\unit{nV}/\sqrt{\unit{Hz}}$]. The ion's motional state with the EIS closed is determined by fitting the amplitude of the Rabi oscillation on the $S_{1/2}$ to $D_{5/2}$ transition \cite{Nagerl2000}, while the state with the EIS open is determined using the sideband amplitude ratio technique \cite{Brownnutt2015}.  These two methods work best in different ranges of motional quanta.  To facilitate plotting both results on the same scale, the linear fit for the EIS-open data is extended, and the experimental data are presented on a smaller scale in an inset.}
\label{fig:heatingRateNoise}
\end{figure}

Voltage noise on the trap electrodes heats the average motional state, $\bar{n}$, of the ion in the trap potential reducing the fidelity of quantum logic gates between ion qubits \cite{Turchette2000, Kirchmair2009}.  We can quantify this effect using standard techniques for measuring heating rates of trapped ions \cite{Brownnutt2015}.  The relationship between voltage noise power and the heating rate of the motional mode, $\dot{\bar{n}}$, is given by:
\begin{equation}
\dot{\bar{n}}(\omega_t) = \frac{q^2}{4m\hbar\omega_t}\frac{S_V(\omega_t)}{D_{\mathrm{eff}}^2},
\label{eq:heatingRate}
\end{equation}
where $m$ and $q$ are the mass and charge of the ion, $\omega_t$ is the (angular) trap frequency, $S_V$ is the square of the voltage noise spectral density and $D_{\mathrm{eff}}$ is a geometric factor determined by the distance from the ion to the trap and the shape of the control electrodes \cite{Leibrandt2007}.  We calculate the value of $D_\mathrm{eff}$ using a finite-element electrostatic simulation to determine the electric field at the ion's position for a given voltage applied to the trap electrodes.  From measurements of $\dot{\bar{n}}$ with the EIS open and closed, we can determine the voltage noise, and thus quantify the isolation of the EIS.

In \figRef{fig:heatingRateNoise}, we present the average motional state of the ion as a function of delay after resolved-sideband cooling, from which we determine the heating rate.  With the EIS closed, we can compare the heating rate and voltage noise, derived via \eqnRef{eq:heatingRate}, with the bench measurement in \figRef{fig:benchNoise}b, but we must account for additional filtering seen by the ion.  The wire bonds that we use to monitor DAC voltages put the trap electrodes in parallel with extra capacitance from a filter board ($C_\mathrm{||}=1\unit{nF}$).  The filter capacitor alone makes a pole at $48\unit{kHz}$ when the switch is closed ($R_\mathrm{EIS}=3.3\unit{k\Omega}$) and reduces the expected voltage noise from $0.98\unit{\mu V}/\sqrt{\unit{Hz}}$ to $32\unit{nV}/\sqrt{\unit{Hz}}$ at $1.5\unit{MHz}$.  This is in agreement with the voltage noise obtained from the heating rate in \figRef{fig:heatingRateNoise}, $1090\pm20\unit{quanta}/\unit{s}\to18.2\pm0.2\unit{nV}/\sqrt{\unit{Hz}}$, within a factor of two.  With the EIS open, the voltage noise is below the noise floor of our spectrum analyzer (\figRef{fig:benchNoise}b), so we expect a heating rate of $180\unit{quanta}/\unit{s}$ or lower.  The measured heating rate with the EIS open, $120\pm30\unit{quanta}/\unit{s}$, is below this limit, but the noise is not reduced by as much as expected given the high isolation of the EIS at lower frequencies.

To further investigate the limitations to the heating rate with the EIS open, we attached a temperature-sensing diode to the surface of the chip.  We found that the temperature of the chip increases from $4\unit{K}$ to just above $50\unit{K}$ when powered on in the CPGA mount.  This temperature increase is due to the power dissipation of the chip-integrated DACs ($500\unit{mW}$), the limited cooling power of the cryostat, and the thermal resistance of the CPGA heatsink \footnote{In an upcoming revision of the chip, we have included a power-down feature which can reduce the power consumption to $16\unit{mW}$ after trap electrodes have charged up and the EIS is opened.}.  At $50\unit{K}$, we expect anomalous electric field noise, arising from the surface of the electrode metal, rather than technical noise in the voltage sources, to limit the heating rate to $30$--$130\unit{quanta}/\unit{s}$ \cite{Brownnutt2015, Bruzewicz2015}, which is in agreement with our measurement in this work.


\section{Output Bandwidth and Filtering}

Finally, we consider the maximum update rate of the voltage sources.  Serialization of input data comes with a speed tradeoff, since trap voltages may only be updated after sending 12 bits to each of the 16 DAC channels.  From SPICE simulation \cite{Nagel1973}, we determine the highest data rate of the SPI bus to be ${\sim}\unit{GHz}$ (maximum voltage update rate ${\sim}10\unit{MHz}$, due to the length of the 192-bit input string); however, in our experiment, we have achieved SPI data rates of $50\unit{MHz}$ (voltage update rate of $250\unit{kHz}$).  Above this speed, there is a sharp decrease in the fidelity of the digital-to-analog conversion.  We presume this limit comes from reflections in the data lines into the vacuum chamber, since this has previously been a source of error in our experience with this SPI bus design.  Using controlled-impedance lines and terminations at the chip would help to reduce these reflections.  

In practice, any voltage updates on the trap electrodes are ultimately limited by the pole at $48\unit{kHz}$ created by the ``on'' resistance of the EIS ($R_\mathrm{EIS}=3.3\unit{k\Omega}$) and the capacitance on the filter board ($C_\mathrm{||}=1\unit{nF}$).  The relatively large resistance of the EIS is not fundamental, and can be reduced in subsequent design revisions.  We could also improve this bandwidth by reducing the parallel filter capacitance, $C_\mathrm{||}$, but this requires careful consideration.  This capacitor also attenuates the Johnson noise associated with the large (${\sim}\unit{T\Omega}$) ``off'' resistance of the EIS; however, at relevant ion frequencies around $1\unit{MHz}$, for parallel capacitance of even a few $\unit{aF}$, this noise source is much smaller than that due to the resistance of the electrode metal and the equivalent series resistance of the capacitor.  More significantly, the additional capacitance to ground shunts leakage from the RF electrode onto the control electrode surfaces by forming a capacitive voltage divider with the effective capacitance between the electrodes \cite{Doret2012}.  From electrostatic simulation, we find that approximately $10\%$ of the RF voltage, if present on the other trap electrodes, can significantly reduce the depth of the potential well seen by the ion.  Acting conservatively, we might choose to keep the RF leakage below $1\%$. The simulated  capacitance between the RF and DC electrodes is between $0.1$ and $0.5\unit{pF}$, so this requires a shunting capacitance of $10$ to $50\unit{pF}$.  With these values, we would expect the bandwidth with the EIS closed to be $1$-$5\unit{MHz}$ and the isolation with the EIS open to be ${>}160\unit{dB}$, which should suppress the DAC technical noise to the same level as the Johnson noise due to the resistance of the trap metal, far below the level of electric field noise due to anomalous heating for traps of this size \cite{Brownnutt2015}. On our chip, a $10\unit{pF}$ capacitor could be integrated in the lower metal layers for each of the trap electrodes without increasing the total device area.


\section{Conclusion}

We have demonstrated the operation of the first ion trap with electrode-control voltage sources embedded into the trap substrate. With the addition of the electrode isolation switch, these sources approach the noise level of standard voltage sources.  The development of these DACs is an important milestone on the way to realizing more complex trapped-ion-system architectures, and our characterization of their performance will be key to assessing their potential for future quantum information processing applications.


\section{Acknowledgements}

We thank Hernan Castro and George Fitch for layout assistance, Peter Murphy, Chris Thoummaraj, and Lee Mailhiot for assistance with chip packaging, and Terry Weir and Gerry Holland for assistance with the bench test setup. This work was sponsored by the Assistant Secretary of Defense for Research and Engineering under Air Force contract number FA8721-05-C-0002. Opinions, interpretations, conclusions, and recommendations are those of the authors and are not necessarily endorsed by the United States Government.


\bibliography{ChipIntegratedSourcesBiblio}

\end{document}